\documentclass[%
  reprint,
  superscriptaddress,
  twocolumn,
  amsmath,
  amssymb,
  10pt,
  aip,
  jcp,
  citeautoscript,
  notitlepage,
]{revtex4-2}

\usepackage[utf8]{inputenc}
\usepackage[T1]{fontenc}
\usepackage{xcolor}
\definecolor{blue}{RGB}{0,0,153}
\usepackage[%
  colorlinks=true,
  allcolors=blue
]{hyperref}
\usepackage{url}
\usepackage{enumitem}
\usepackage{amsfonts}
\usepackage{amssymb}
\usepackage{amsmath}
\usepackage{mathtools}
\usepackage[ruled,vlined]{algorithm2e}
\usepackage{lmodern}
\usepackage[normalem]{ulem}

\DeclarePairedDelimiter\ppar{(}{)}              
\DeclarePairedDelimiter\pnrm{\lVert}{\rVert}    
\DeclarePairedDelimiter\pset{\{}{\}}            

\newcommand{\rtab}[1]{Tab.~\ref{#1}}
\newcommand{\rfig}[1]{Fig.~\ref{#1}}
\newcommand{\rsfig}[2]{Fig.~\ref{#1}(#2)}

\newcommand{\rref}[1]{Ref.~\citenum{#1}}
\newcommand{\req}[1]{Eq.~\ref{#1}}

\newcommand{\dd}[1]{\operatorname{d#1}}

\newcommand{\bz}{\mathbf{z}}
\newcommand{\bx}{\mathbf{x}}

\newcommand{\dx}{\dd{\mathbf{x}}}
\newcommand{\e}{\operatorname{e}}
\newcommand{\kT}{k_{\mathrm{B}}T}

\newcommand{\chg}[1]{{\color{black}{#1}}}

\begin{document}
\title{%
  Learning Markovian Dynamics with Spectral Maps
}

\author{Jakub Rydzewski}
\email[]{jr@fizyka.umk.pl}
\author{Tuğçe Gökdemir}
\affiliation{%
  Institute of Physics,
  Faculty of Physics, Astronomy and Informatics,
  Nicolaus Copernicus University,
  Grudziadzka 5, 87-100 Toru\'n, Poland
}

\date{\today}

\begin{abstract}
The long-time behavior of many complex molecular systems can often be described by Markovian dynamics in a slow subspace spanned by a few reaction coordinates referred to as collective variables (CVs). However, determining CVs poses a fundamental challenge in chemical physics. Depending on intuition or trial and error to construct CVs can lead to non-Markovian dynamics with long memory effects, hindering analysis. To address this problem, we continue to develop a recently introduced deep-learning technique called spectral map \href{https://doi.org/10.1021/acs.jpclett.3c01101}{[Rydzewski, {\it J. Phys. Chem. Lett.} {\bf 2023}, 14, 22, 5216--5220]}. Spectral map learns slow CVs by maximizing a spectral gap of a Markov transition matrix describing anisotropic diffusion. Here, to represent heterogeneous and multiscale free-energy landscapes with spectral map, we implement an adaptive algorithm to estimate transition probabilities. Through a Markov state model analysis, we validate that spectral map learns slow CVs related to the dominant relaxation timescales and discerns between long-lived metastable states.
\end{abstract}

\maketitle

\section{Introduction}
Understanding the long timescale dynamics of complex molecular systems constitutes a fundamental problem in chemical physics~\cite{dfrenkel:mc}. This dynamics is often governed by rare transitions between multiple long-lived metastable states that involve overcoming barriers much higher than the thermal energy ($\gg \kT$)~\cite{valsson2016enhancing,yang2019enhanced}. Although having many spatial and temporal scales, systems exhibiting metastable characteristics frequently display slow relaxation dynamics that can be captured by a few reaction coordinates, also called collective variables (CVs), to demarcate between physical states~\cite{rohrdanz2013discovering,noe2017collective}. The long timescale dynamics is often considered effectively Markovian and modeled as diffusion in a free-energy landscape spanned by CVs~\cite{berezhkovskii2005one,berezhkovskii2011time,maragliano2006string}. Under this view, the dynamics mainly depends on slowly varying variables, while fast degrees of freedom are treated as uncoupled thermal noise, leading to adiabatic timescale separation~\cite{zwanzig2001nonequilibrium}.

Relying merely on physical intuition or trial and error to identify slow CVs can be unsystematic and result in highly non-Markovian dynamics with long memory effects, which can hinder our understanding of the underlying physical process~\cite{lange2006collective,micheletti2008optimal}. The problem of timescale separation has been addressed in the Mori--Zwanzig projection operator formalism by constructing a generalized Langevin equation that describes the dynamics in a slow subspace. However, as noted by Zwanzig, ``no a priori choice is specifically indicated by theory''~\cite{zwanzig1961memory} to select this slow subspace. Recently, there has been considerable interest in constructing reduced representations of complex systems using data-driven techniques~\cite{ceriotti2019unsupervised,wang2020machine,chen2023chasing,rydzewski2023manifold}. Many such techniques are developed to learn CVs by separating slow and fast kinetics, which is particularly useful for metastable systems with multiple temporal scales~\cite{coifman2005geometric,singer2009detecting,naritomi2011slow,ferguson2011integrating,rohrdanz2011determination,hernandez2013identification,tiwary2016spectral,wu2017variational,chiavazzo2017intrinsic,wehmeyer2018time,thiede2019galerkin,chen2019nonlinear,bonati2021deep,morishita2021time,evans2022computing,chen2023discovering,rydzewski2023selecting,rydzewski2023spectral}.

In this Communication, we further improve spectral map~\cite{rydzewski2023spectral}, an unsupervised statistical learning technique inspired by diffusion map~\cite{nadler2006diffusion,coifman2008diffusion,singer2008non} and parametric dimensionality reduction~\cite{hinton2006reducing,maaten2009learning,zhang2018unfolding,rydzewski2021multiscale,rydzewski2022reweighted}. Spectral map constructs slow CVs to describe Markovian dynamics by maximizing the spectral gap between slow and fast eigenvalues of a Markov transition matrix, increasing timescale separation, and preventing memory effects. We improve the learning process of spectral map by implementing an adaptive algorithm to estimate transition probabilities, enabling the representation of multiscale and heterogeneous free-energy landscapes with long-lived metastable states according to their physical characteristics. Finally, through a standard Markov state model analysis, we validate that CVs constructed by spectral map encode dominant slow relaxation timescales.

\section{Collective Variables}
Consider a high-dimensional system described by $n$ configuration variables $\bx = (x_1, \dots, x_n)$ whose dynamics at temperature $T$ and a potential energy function $U(\bx)$ is sampled from an unknown equilibrium distribution. If we represent the system by its microscopic coordinates, the dynamics follows a canonical equilibrium distribution given by the Boltzmann density $p(\bx) = \e^{-\beta U(\bx)}/Z$, where $\beta=1/(k_{\mathrm{B}}T)$ is the inverse temperature, $k_{\mathrm{B}}$ is the Boltzmann constant and $Z=\int\dx\e^{-\beta U(\bx)}$ is the partition function of the system.

We simplify the high-dimensional configuration space by mapping it into a reduced space $\bz=\ppar*{z_1, \dots, z_d}$ given by a set of $d$ functions of the configuration variables, commonly referred to as CVs, where $d \ll n$. We encapsulate these functions in a target mapping~\cite{rydzewski2021multiscale,rydzewski2022reweighted,rydzewski2023manifold}:
\begin{equation}
  \label{eq:target-mapping}
  \bz = \xi_{w}(\bx) \equiv \pset[\big]{\xi_k(\bx, w)}_{k=1}^d,
\end{equation}
where $w$ are learnable parameters ensuring that the target mapping describes slow dynamics and minimizes memory effects. By sampling the system in the CV space, its dynamics follows a marginal equilibrium density $p(\bz)\propto\e^{-\beta F(\bz)}$, where $F(\bz)$ is a free-energy landscape:
\begin{equation}
  \label{eq:fel}
  F(\bz) = -\frac{1}{\beta}\log\int\dx\,\delta\ppar*{\bz - \xi_w(\bx)}\e^{-\beta U(\bx)}
\end{equation}
defined modulo an irrelevant constant.

\begin{figure*}[t]
  \includegraphics{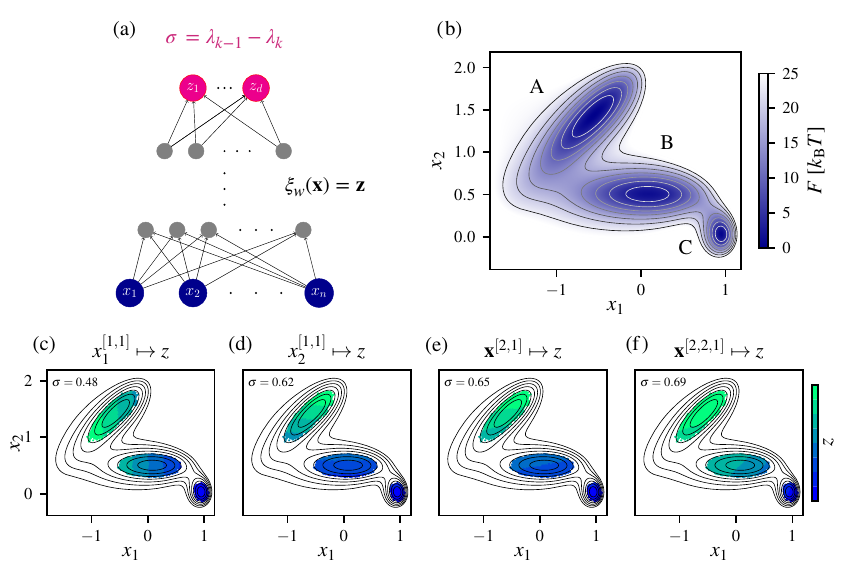}
  \caption{Training the target mapping $\xi_w(\bx)=\bz$ for the three-state M\"uller--Brown potential from \rref{bonati2023unified}. (a) Schematic outline of a neural network used to model the target mapping. The reduction from high-dimensional variables $\bx$ is performed to obtain low-dimensional slow CVs $\bz$ by maximizing the spectral gap $\sigma=\lambda_{k-1}-\lambda_k$, where $k$ is the number of the metastable states in the CV space, and $\pset{\lambda_k}$ are eigenvalues from the spectral decomposition of the Markov transition matrix $M$ (\req{eq:markov}). (b) Free-energy landscape of a particle moving in the two-dimensional space $\bx=(x_1,x_2)$ with barriers between the states of around 20 $\kT$ and a kinetic bottleneck between states B and C. (c-f) Examples of training the target mappings with different network architectures, e.g., $\bx^{[a,\dots,b]}\mapsto z$ denotes mapping the $\bx$ variable through a network consisting of $[a,\dots,b]$ layers to the $z$ CV. Spectral gap values $\sigma$ are given in the top-left corners. The following variables are used as CVs: (c) the $x_1$ variable, (d) the $x_2$ variable, (e) a linear combination of $x_1$ and $x_2$, (f) and a nonlinear combination of $x_1$ and $x_2$. For additional details and results, see the Supplementary Material (Sec. S3).}
  \label{fig:mb}
\end{figure*}

\section{Spectral Map}
To estimate the separation between effective timescales characteristic of the system, we model its reduced dynamics as a Markov chain using kernel functions. To measure the similarity between CV samples $\bz_k$ and $\bz_l$, we introduce a Gaussian kernel for the reduced representation:
\begin{equation}
  \label{eq:gauss}
  G(\bz_k,\bz_l)=\exp\ppar*{-\frac{1}{\varepsilon}{\pnrm{\bz_k-\bz_l}^2}}
\end{equation}
where $\pnrm{\bz_k-\bz_l}$ denotes pairwise Euclidean distances between every pair of CV samples $k,l=1,\dots,N$ and $N$ is the number of samples. The Gaussian kernel exhibits a notion of locality by defining a neighborhood around each sample of radius $\varepsilon$~\cite{coifman2005geometric}.

In our previous proof-of-concept work that proposed spectral map~\cite{rydzewski2023spectral}, $\varepsilon$ was kept constant throughout the learning process. However, metastable states often have multimodal characteristics, resulting in spatially heterogeneous free-energy landscapes. Therefore, to improve the ability of spectral map to adjust to metastable states, we estimate the sample-dependent scale factors by adaptively balancing local and global scales as:
\begin{equation}
  \label{eq:scale}
  \varepsilon_{kl}(r)=\pnrm{\bz_k - \eta_r(\bz_k)} \cdot \pnrm{\bz_l - \eta_r(\bz_l)},
\end{equation}
where each term defines a ball centered at $\bz$ of radius $\eta_r(\bz)>0$. For convenience, we define the radius by the fraction of the neighborhood size $r\in[0,1]$, allowing us to decide which scale is more relevant. Specifically, the Gaussian kernel describes a local neighborhood around each sample for values $r$ close to 0 (i.e., the nearest neighbors), which correspond to deep and narrow states. For values of $r$ around 1 (the farthest neighbors), it considers more global information, corresponding to shallow and wide states. Intermediate values of $r$ maintain a balance between spatial scales. For additional details about estimating $\varepsilon_{kl}$, see the Supplementary Material (Sec. S2).

As the marginal equilibrium density in the CV space is often far from uniform for dynamical systems with complex free-energy landscapes, we need a density-preserving kernel for data sampled from any underlying probability distribution. For this, we employ an anisotropic diffusion kernel as introduced for diffusion maps~\cite{nadler2006diffusion}:
\begin{equation}
  \label{eq:kernel}
  K(\bz_k,\bz_l) = \frac{G(\bz_k,\bz_l)}{\sqrt{\varrho(\bz_k)\varrho(\bz_l)}},
\end{equation}
where $\varrho(\bz_k)=\sum_l G(\bz_k,\bz_l)$ is a kernel density estimate. Next, we build a Markov transition matrix by row-normalizing $K$:
\begin{equation}
  m_{kl} \sim M(\bz_k,\bz_l) = \frac{K(\bz_k,\bz_l)}{\sum_n K(\bz_k,\bz_n)}
  \label{eq:markov}
\end{equation}
which models a discrete Markov chain in the CV space $m_{kl}=\mathrm{Pr}\pset{\bz_{\tau+1}=\bz_l\;|\;\bz_\tau=\bz_k}$ expressing a probability of transition between CV samples $\bz_k$ and $\bz_l$ in an auxiliary (non-physical) time $\tau$. The Markov chain approximates the long-time asymptotics of the system by describing the dynamics by the Fokker--Planck anisotropic diffusion~\cite{nadler2006diffusion}.

\begin{figure*}[t]
  \includegraphics{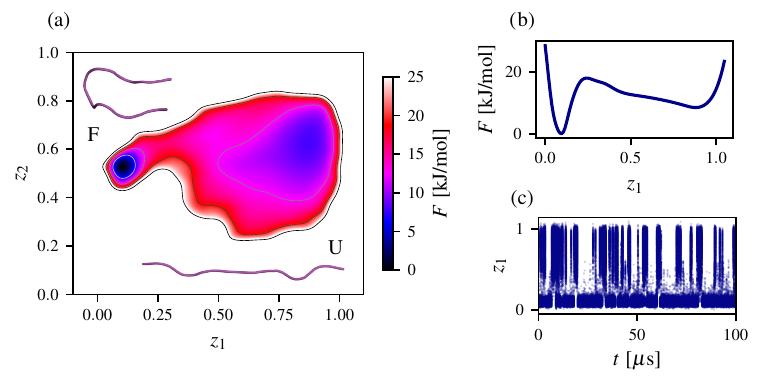}
  \caption{Learning slow CVs for the reversible folding of CLN025 in solvent sampled through a 100-$\mu$ molecular dynamics simulation. The dataset is obtained from \rref{lindorff2011fast}. (a) Spectral map and the corresponding free-energy landscape $F$ of CLN025 showing two distinct free-energy basins: the main and well-defined folded (F) and more loosely structured unfolded (U) metastable states, separated by a free-energy barrier of around 20 kJ/mol. Representative conformations from the metastable states are shown. The aspect ratio of axes is preserved. (b) Free-energy profile shown as a function of the $z_1$ CV. (c) Time series for the $z_1$ CV showing transitions between the states.}
  \label{fig:spectral-map}
\end{figure*}

Finally, to estimate the dominant timescales encoded in the system, we perform a spectral decomposition of the Markov transition matrix:
\begin{equation}
  M\Psi_k=\lambda_k\Psi_k,  
\end{equation}
where $\Psi_k$ and $\lambda_k$ are the $k$-th right eigenfunctions and eigenvalues of $M$, respectively. The real-valued eigenvalues of $M$ are (sorted in non-ascending order):
\begin{equation}
  \lambda_0 = 1 > \lambda_1 \dots \ge \lambda_N,
\end{equation}
where the eigenvalue $\lambda_0$ corresponds to the equilibrium distribution of the Markov chain (\req{eq:markov}) given by the eigenfunction $\Psi_0$. The dominant eigenvalues related to the slowest relaxation timescales in the system can be found by associating each eigenvalue with an effective timescale~\cite{bovier2002metastability}:
\begin{equation}
  t_k = -\frac{1}{\log \lambda_k}.
  \label{eq:timescale}
\end{equation}
The largest gap between neighboring eigenvalues is called the spectral gap and determines the degree of the timescale separation between the slow and fast processes:
\begin{equation}
  \label{eq:spectral-gap}
  \sigma = \lambda_{k-1} - \lambda_k,
\end{equation}
where $k > 0$ indicates the number of metastable states in the CV space~\cite{gaveau1998theory}. 

\chg{
The theory of spectral characterization of metastable states (see works by Gaveau and Schulman~\cite{gaveau1996master,gaveau1998theory} and references therein) explains that the spectral gap and degree of degeneracy in eigenvalue spectrum are related to the timescale separation and Markovian dynamics. If the eigenvalue of the Markov transition matrix is nearly degenerate $k+1$ times, it indicates that the equilibrium distribution breaks into $k$ metastable states with infrequent transitions between them. The converse is also true: if the equilibrium density breaks into metastable states separated by a free-energy barrier much larger than the thermal energy $\kT$, there is eigenvalue degeneracy. 

To achieve the reduced dynamics that is effectively Markovian, it is crucial to have a spectral gap between neighboring eigenvalues, along with the near degeneracy of the dominant eigenvalue. This condition is essential for the maximal spectral gap at $k$ to lead to the separation into $k$ metastable states, which can help reduce memory effects in the dynamics. Therefore, we consider the spectral gap as a scoring function in spectral map.
}

\section{Applications}

\subsection{M\"uller--Brown Potential}
As a simple demonstration, we apply spectral map to a dataset consisting of trajectories of a particle moving in a modified three-state M\"uller--Brown potential [\rsfig{fig:mb}{b}]. Following \rref{bonati2023unified}, we modify the standard M\"uller--Brown potential by adding a kinetic bottleneck between states B and C that is mainly responsible for the slowest timescale in the dynamics (see Sec. S3 in the Supplementary Material). The input variables for the target mappings are the $\bx=(x_1,x_2)$ coordinates of the particle. We use the Adam optimizer with a learning rate of $10^{-3}$. The dataset consisting of 80 trajectories initiated at random from the three metastable states is divided into batches of size 2000. The training for $k=3$ is carried through 30 epochs. The trajectories are generated by a Langevin integrator as implemented in PLUMED~\cite{plumed,plumed-nest}, using a temperature of 1, a friction coefficient of 10, and a time step of 0.005. We set the fraction of neighborhood size used to estimate sample-dependent scale factors to $r=0.5$ (\req{eq:scale}).

In \rsfig{fig:mb}{c}, we can see that if the $x_1$ variable is used as a CV, the spectral gap is just around $\sigma=0.48$. Treating the $x_2$ as a CV results in an improved spectral gap of 0.62, as $x_2$ is slower than $x_1$ and better separates the three states [\rsfig{fig:mb}{d}]. Note that the CVs shown in \rfig{fig:mb}(c-d) are identity mappings and cannot be maximized. They are used as a baseline to calculate the reference values of the spectral gap. Using a linear combination of $x_1$ and $x_2$, we can slightly increase the separation between the states [\rsfig{fig:mb}{e}]. By adding a hidden layer with the ELU activation function to the target mapping [\rsfig{fig:mb}{f}], we obtain a separation with the main states A and B closer to each other. State C is pushed farther away as the particle reaching this state must overcome the kinetic bottleneck. This clearly shows that the nonlinear CV captures this slowest transition in the dynamics. As the transitions between states A and B are faster, these states are relatively closer to each other in the reduced representation. For additional details and results, see the Supplementary Material (Sec. S3).

\subsection{Chignolin}
As the next application of our method, we consider the folding process of a ten-residue protein chignolin (CLN025) in solvent. A 100-$\mu$s unbiased molecular dynamics simulation of CLN025 at its melting temperature of 340 K is obtained from \rref{lindorff2011fast}. The folding (also termed ``zipping'') of CLN025 is of interest due to its $\beta$-hairpin structure and can serve as a building block to understand more complex processes.

As a high-dimensional representation, we use pairwise Euclidean distances between C$\alpha$ atoms, amounting to $n=45$ configuration variables. \chg{The dataset consists of 10,000 samples extracted from the simulation every 10 ns and randomly split into the training and validation sets of sizes 6000 and 4000, respectively.} The training of the target mapping is carried out using $k=2$ for 100 epochs with data batches of 2000 samples. Once the target mapping is trained, we evaluate all samples from the trajectory (every 200 ps) to construct a free-energy landscape. To model the target mapping, we use a neural network of size [45, 200, 100, 50, 2], with each hidden layer employing the ELU activation function. We use the Adam optimizer with a learning rate of $10^{-3}$. We set the fraction of neighborhood size used to estimate sample-dependent scale factors to $r=0.65$ by exploring which value of $r$ corresponds to the largest spectral gap (see Sec. S3 in the Supplementary Material for more details).

We present our results in \rfig{fig:spectral-map}. We can see that the free-energy landscape spanned by the slow CVs identified by spectral map depicts the folded and unfolded metastable states of CLN025 and rare transitions between these states by separating them with a barrier of around 20 kJ/mol. An interesting feature of these CVs is that the shape of the metastable states is preserved by the adaptive algorithm for computing sample-dependent scale factors. Namely, the folded state is well-defined, which is shown by its small size and the deepest free-energy minimum. In contrast, as the unfolded conformations are more loosely constrained, they are in a wider and shallower free-energy basin. Remarkably, the shape and depths of the metastable states and the height of the free-energy barrier are almost indistinguishable from results presented in \rref{lindorff2011fast}.

In comparison to CVs calculated for CLN025 in our previous proof-of-concept work that introduced spectral map, where a constant scale factor is used to construct the Markov transition matrix (see \rref{rydzewski2023spectral}), it is especially evident that the mapping of the unfolded state improved. Here, it resembles more accurately the physical characteristics of an ensemble comprising many degenerate configurations with a limited number of native contacts~\cite{best2013native}. Therefore, we can see that estimating a neighborhood adaptively, instead of providing a constant value, enables a physically meaningful reconstruction of the free-energy landscape. 

\chg{Furthermore, we use the spectral gap and sum of eigenvalues to evaluate the quality of two-dimensional projections of the CLN025 dataset computed using time-lagged independent component analysis~\cite{hernandez2013identification} for several values of lag times. Comparing these results to the scores calculated from the CV space constructed with spectral map indicates that the CVs found with spectral map are characterized by both larger values of the spectral gap and the sum of eigenvalues. For more details, we refer to Sec. S3 in the Supplementary Material.}

To validate that the learned CVs correspond to the slowest relaxation process in the dynamics of CLN025 and do not mix the folded and unfolded states, we perform a standard Markov state model analysis~\cite{prinz2011markov}. The quality of a Markov state model and the estimation of relaxation times can depend heavily on the reaction coordinate describing the studied physical process~\cite{mckiernan2017modeling}. For example, using non-Markovian CVs that do not distinguish between long-lived states to build a Markov state model may lead to an accumulation of errors in the next steps of modeling and, ultimately, in estimated kinetics. However, using a reaction coordinate that accurately represents slow kinetics can significantly improve the accuracy of a Markov state model. Therefore, to validate that the CVs provided by spectral map accurately describe the slow modes of CLN025, we need to obtain relaxation timescale estimates that match reference values~\cite{lindorff2011fast}.

\begin{table}
  \caption{Folding (F) and unfolding (U) times of CLN025. Mean times and standard deviations are estimated by constructing a Markov state model from the slow CVs learned by spectral map. Reference values are taken from \rref{lindorff2011fast}.}
  \begin{tabular}{lccc}
    \hline
    Process & Mean [$\mu$s] & St. Dev. [$\mu$s] & Ref. [$\mu$s]~\cite{lindorff2011fast} \\
    \hline
    Folding   & 0.58 & 0.01 & 0.61 \\
    Unfolding & 2.08 & 0.02 & 2.24 \\
    \hline
  \end{tabular}
  \label{tab:times}
\end{table}

Using the standard pipeline for constructing a Markov state model (see Sec. S4 in the Supplementary Material for details), we find that there is a single dominant relaxation process encoded in the CV learned by spectral map that corresponds to the reversible folding of CLN025. Other processes occur on a timescale below a lag time $\tau$ of 100 ns. We observe a plateau in an implied timescale convergence analysis for $>100$ ns, showing that the model is Markovian. Our analysis shows the model is not sensitive to changes in its parameters, as indicated by the Chapman--Kolmogorov test (performed up to 1 $\mu$s). Having verified that the resulting Markov state model is very accurate, we estimate folding and unfolding times. The means and standard deviations of the folding and unfolding of CLN025 are shown in \rtab{tab:times}. Our estimates agree remarkably well compared to times reported by Lindorf-Larssen et al., who used a native contact-based definition of the folded and unfolded states~\cite{lindorff2011fast}. Overall, our results indicate that spectral map uncovers slow CVs correctly by ensuring that the selected number of long-lived metastable states and rare transitions between them are represented in the reduced representation.

Overall, we have shown that the learned slow CVs separate the folded and unfolded states of CLN025 according to the slowest relaxation timescale, with faster processes being negligible. Furthermore, we have validated that the learned CVs are correct by performing a standard Markov state model analysis. The slow CVs, learned by spectral map without any data preprocessing and landmark subsampling~\cite{rydzewski2023manifold}, have allowed us to build a high-quality Markov state model and, thus, to estimate the relaxation timescales of CLN025 accurately. Moreover, supplementing the Markov state model with the slow reaction coordinates has simplified parameter search in the model, as shown by little dependence on selected parameters (see Sec. S4 in the Supplementary Material). This result is especially promising as selecting parameters in Markov state models is known to be difficult in some cases~\cite{nedialkova2014diffusion,mcgibbon2017identification,husic2017note,suarez2021markov}.

\section{Conclusions}
Many techniques for learning slow CVs rely partly on identifying lag times. In contrast, spectral map learns CVs describing the slowest modes and dominant relaxation times without requiring a specific lag time. This is possible as spectral map estimates transition probabilities between samples, unlike other techniques that rely on counting configurations. Selecting an incorrect lag time can lead to non-Markovian dynamics and result in non-negligible memory effects, which can mix slow and fast timescales, causing significant errors in estimated kinetics~\cite{schultze2021time,kozlowski2023uncertainties}. However, it is important to note that spectral map exchanges temporal parameters (lag time) for spatial parameters (neighborhood size). Estimating scale factors is also known to affect the reduced representation~\cite{berry2015nonparametric,berry2016variable}. Although methods for identifying slow CVs using a lag time~\cite{hernandez2013identification,wehmeyer2018time,mardt2018vampnets} are well established and commonly used, spectral map is a recently developed technique. Therefore, it is yet to be determined if estimating scale factors is more practical than selecting lag times.

Spectral map employs the anisotropic diffusion kernel to estimate transition probabilities and maximize the timescale separation. This kernel has been first introduced for diffusion map~\cite{coifman2005geometric}, and therefore, it is important to compare these two techniques. In diffusion maps, the reduced space is approximated using eigenfunctions of a Fokker--Planck operator (diffusion coordinates) constructed from samples in the configuration space. Because of that, the validity of the eigenfunctions cannot be self-consistently verified, and their quality cannot be improved. Additionally, from a technical perspective, eigendecompositions of large transition matrices can be computationally expensive. In contrast to diffusion map, spectral map performs the spectral decomposition of a Markov transition matrix in the reduced space, which is perhaps the main difference between the two techniques. Spectral map does not rely on eigenfunctions as approximations of the reduced representation and only requires computing eigenvalues. These eigenvalues are then used to maximize the spectral gap to iteratively improve CVs corresponding to slow kinetics through the learning process. Initially, the reduced dynamics can have non-Markovian characteristics. As the spectral gap is maximized, memory effects are reduced. Moreover, the eigendecompositions are computed from data batches, which reduces computational costs.

In this Communication, we have further improved spectral map, an unsupervised machine-learning technique for learning slow CVs describing Markovian dynamics of complex metastable systems. By implementing an adaptive algorithm for estimating transition probabilities, we have added to spectral map the ability to represent heterogeneous and multiscale free-energy landscapes. We have verified with a standard Markov state model analysis that spectral map can be successfully used to construct CVs corresponding to the slowest modes. Our technique is applicable to a variety of physical processes with multiple temporal scales. Although spectral map is still in its preliminary version, it has shown potential in computing slow reaction coordinates and deserves further development.

\section*{Supplementary Material}
Computational details and results, including information about simulation datasets, learning, and Markov state models are available in the Supplementary Material.

\section*{Acknowledgments}
J. R. acknowledges funding from the Polish Science Foundation (START), the Ministry of Science and Higher Education in Poland, and the Japan Society for the Promotion of Science (JSPS). J. R. and T. G. are supported by the National Science Center in Poland (Sonata 2021/43/D/ST4/00920, ``Statistical Learning of Slow Collective Variables from Atomistic Simulations''). D. E. Shaw Research is acknowledged for providing the CLN025 dataset.

\section*{Author Declarations}
\subsection*{Conflict of Interest}
The authors have no conflicts to disclose.

\subsection*{Author Contributions}
\noindent{\bf Jakub Rydzewski:} Conceptualization (leading); Investigation (equal); Methodology (leading); Software (leading); Supervision (leading); Visualization (leading); Writing -- original draft (leading); Writing -- review \& editing (equal). {\bf Tuğçe Gökdemir:} Conceptualization (supporting); Investigation (equal); Methodology (supporting); Software (supporting); Supervision (supporting); Visualization (supporting); Writing -- original draft (supporting); Writing -- review \& editing (equal).

\section*{Data Availability}
Data and a reference implementation of spectral map are available on PLUMED-NEST~\cite{plumed-nest} (\url{www.plumed-nest.org}), the public repository of the PLUMED consortium, as plumID:24.005.

\section*{References}
\bibliography{main.bib}

\end{document}